\documentclass{article}
\usepackage{amssymb}

\usepackage{graphicx}
\usepackage{amsmath}

\input{tcilatex}

\begin{document}

\title{\textbf{Stochastic Self-Similar and Fractal Universe}}
\author{G.Iovane\thanks{%
iovane@diima.unisa.it} \\
Dipartimento di Ingegneria dell'Informazione e Matematica Applicata,\\
Universit\'a di Salerno, Italy. \\
Gruppo Nazionale di Fisica Matematica, Italy\\
E.Laserra\\
Dipartimento di Matematica e Informatica, Universit\'a di Salerno, Italy \\
Gruppo Nazionale di Fisica Matematica, Italy\\
F.S.Tortoriello\\
Universit\'a di Salerno, Italy \\
Gruppo Nazionale di Fisica Matematica, Italy.\\
\\
}
\date{Last changed 20.07.03}
\maketitle

\begin{abstract}
The structures formation of the Universe appears as if it were a classically
self-similar random process at all astrophysical scales. An agreement is
demonstrated for the present hypotheses of segregation with a size of
astrophysical structures by using a comparison between quantum quantities
and astrophysical ones. We present the observed segregated Universe as the
result of a fundamental self-similar law, which generalizes the Compton
wavelength relation. It appears that the Universe has a memory of its
quantum origin as suggested by R.Penrose with respect to quasi-crystal. A
more accurate analysis shows that the present theory can be extended from
the astrophysical to the nuclear scale by using generalized (stochastically)
self-similar random process. This transition is connected to the relevant
presence of the electromagnetic and nuclear interactions inside the matter.
In this sense, the presented rule is correct from a subatomic scale to an
astrophysical one. We discuss the near full agreement at organic cell scale
and human scale too. Consequently the Universe, with its structures at all
scales (atomic nucleus, organic cell, human, planet, solar system, galaxy,
clusters of galaxy, super clusters of galaxy), could have a fundamental
quantum reason. In conclusion, we analyze the spatial dimensions of the
objects in the Universe as well as spacetime dimensions. The result is that
it seems we live in an El Naschie's E infinity Cantorian spacetime; so we
must seriously start considering fractal geometry as the geometry of nature,
a type of arena where the laws of physics appear at each scale in a
self--similar way as advocated long ago by the Swedish school of
astrophysics.
\end{abstract}

%

\section{Introduction}

What is the geometry of the universe? Has the universe a memory of its
quantum and relativistic origin?

In 1965 Sakharov indicated that quantum primordial fluctuations should have
expanded towards the present epoch leading first to classical energy-density
perturbations and, after the decoupling from the cosmological background, to
the observed galaxies, clusters and superclusters of galaxies \cite{Sakharov}%
.

A relevant contribution was given by L.Nottale starting from 1993. In many
papers he extends Einstein's principle of relativity to scale
transformations in the framework of the theory of scale relativity. In
particular, he showed that a continuous but non differentiable space-time is
necessarily fractal \cite{Nottale1}, \cite{Nottale2}, \cite{Nottale3}, \cite
{Nottale4}. In this work, we present a complementary approach starting from
the well--known Random Walk equation or Brownian motion relation that was
firstly used by Eddington \cite{Elnaschie3}, \cite{Sidharth},\cite
{capozziello}. Following this line we arrive at a self-similar universe;
which was firstly considered by the Swedish Astronomers Charlier \cite{Rees}%
. By taking into account a generalization of Compton wavelength rule, the
model realizes a segregated universe, where the sizes of astrophysical
structures can fit the observations (e.g. COBE, IRAS, and surveys of large
scale structures \cite{Lapparent}). The idea, that a rule can exist among
the fundamental constants, was presented by Dirac and by
Eddington--Weinberg, but these rules were exact at Universe scale or
subatomic scale. Here, a scale invariant rule is presented. Thanks to this
relation the Universe appears self similar and its self similarity is
governed by fundamental quantum quantities, like the Plank constant \textit{h%
}, and relativistic constants, like the speed of light \textit{c}.

Actually, there are some theories of gravity which are obtained from the
Einstein-Hilbert gravitational action by adding scalar fields or curvature
invariants of the form $\phi^{2}R$, $R^{2}$, $R_{\mu\nu}R^{\mu\nu}$, $%
R\square R$ \cite{La},\cite{Buch},\cite{Stelle}. However, in the weak-limit
approximation, all these theories fit very well with the experiments of
Einstein's general relativity (tested only in this limit) \cite{Will}.
Moreover, the observations show a structure of Universe with scaling rules,
where we can see globular clusters, single clusters or superclusters of
galaxies, in which stars can be treated as massive point-like constituents
of a universe mad of dust.

Why does the Universe appear with fixed scales, where matter can be
clustered? The right question is not the previous one, but the following
one: does the Universe have quantum nature at all scales? It appears that
the Universe has a memory of its quantum origin like as suggested by
R.Penrose with respect to quasi-crystal \cite{Penrose}. Particularly, it is
related to Penrose tiling and thus to $\varepsilon ^{(\infty )}$ theory
(Cantorian spacetime theory) as proposed by M.S. El Naschie \cite{Elnaschie1}%
,\cite{Elnaschie2} as well as in A.Connes Noncommutative Geometry \cite
{Connes}.

Some remarks are presented about the segregation of the Universe with
respect to an Eddington--Weinberg--like relation ($h=m\sqrt{mGR}$, where $G$
is the Newton gravitational constant, $m$ is the mass of nucleon, $R$ is the
radius of the Universe and $h$ is the quantum unit of action). In
particular, we analyze the scale invariant law $R(N)=\frac{h}{Mc}N^{\alpha }$%
, where $R$ is the radius of the astrophysical structures, $h$ is the Planck
constant, $M$ is the total Mass of the self-gravitating system, $c$ the
speed of light, $N$ the number of nucleons into the structures and $\alpha
\simeq 3/2$. This relation is the Compton wavelength for $N=1$. The Newton
gravitational constant $G$ probably plays no fundamental role in respect to
the dimension of an object, while it becomes relevant in the interaction
between the objects. So it is obvious that we have not found $G$ in the
constitutive relations.

Another relevant point is the connection of the presented law with the
Golden Mean. From the art to the science the role of the Golden Mean is well
known \cite{Cook}. Here our expression agree with the Golden Mean and with
the gross law of Fibonacci and Lucas \cite{Vajda}.

The paper is organized as follows: we find the astrophysical scenario in
Sec.2; Sec.3 presents a short review of definitions and properties for
classic and stochastic self-similar random processes; Sec.4 is devoted to
studying the exact determination of the power law at all significant scales
and not only at astrophysical scales; in Sec.5 we briefly analyze some
fundamental consequences from physical and geometric points of view and
finally conclusions are drawn in Sec.6.

\section{Astrophysical scenario: quantum fluctuations and size of
self-gravitating systems}

As it is known luminous matter appears segregated at different scale; in
particular, we can distinguish among globular clusters, galaxies, clusters
and superclusters of galaxies through their spatial dimensions \cite{Binney}%
, \cite{Vorontsov}. Table 1 recalls the dimensions and masses of previous
systems \cite{Abell}, \cite{Peebles}. 
\begin{eqnarray*}
&&\frame{$
\begin{array}{lll}
\text{\textbf{System Type}} & \text{\textbf{Length}} & \text{\textbf{Mass(}}%
M_{\odot }\text{\textbf{)}} \\ 
\text{Globular Clusters} & R_{GC}\thicksim 10pc & M_{GC}\thicksim 10^{6\div
7} \\ 
\text{Galaxies} & R_{G}\thicksim 1\div 10kpc & M_{G}\thicksim 10^{10\div 12}
\\ 
\text{Cluster of galaxies} & R_{CG}\thicksim 1.5h^{-1}Mpc & M_{CG}\thicksim
10^{15}h^{-1} \\ 
\text{Supercluster of galaxies} & R_{SCG}\thicksim 10\div 100h^{-1}Mpc & 
M_{SCG}\thicksim 10^{15\div 17}h^{-1}
\end{array}
$} \\
&&\text{Table 1: Classification of astrophysical systems by length and mass, 
} \\
&&\text{where h is the dimensionless Hubble constant whose value is in the}
\\
&&\text{ range [0.5,1].}
\end{eqnarray*}
In this paragraph, we consider systems where gravity is the only interaction
among the constituents. For this reason, stars or objects smaller or larger
than stars, where electromagnetic or nuclear interaction could be relevant,
are not taken into account. Moreover, in the work only luminous matter is
considered (dark matter will be considered in a future paper). Under these
hypotheses, we can see stars as granular constituents of dust globular
clusters or galaxies and so on. Moreover, a typical interaction length can
be defined as a quantity which is proportional to the size of the system
which contains the constituents. In other words, for each system we consider
a maximum length, corresponding to its size, that plays the same role as the
interaction length.

In 1965 Sakharov argued that quantum primordial fluctuations had to be
related to cosmological evolution and to the dynamics of astrophysical
systems \cite{Sakharov}. Eddington and later on Weinberg wrote the relevant
relationship between quantum quantities and the cosmological ones: 
\begin{equation}
h\cong G^{1/2}m^{3/2}R^{1/2},
\end{equation}
where $h$ is the Plank constant, $G$ is the gravitational constant, $m$ is
the mass of nucleon, and $R$ is the radius of Universe.

By following the Eddington-Weinberg (E-W) approach, we can write a general
relationship between the radius R of the self-gravitating system and its
number of nucleons. While the E-W relationship was written only for the
radius of Universe, we present a relationship which is scale invariant, so
it can be adopted for all types of self-gravitating systems (and also for
the entire universe).

It is interesting to note that if we write: 
\begin{equation}
R(N)=\frac{h}{Mc}N^{\alpha },
\end{equation}
with $\alpha =3/2,$ for $M=M_{G}\thicksim 10^{10\div 12}M_{\odot }$ and $%
N=10^{68}$(this is approximately the number of nucleons in a galaxy), we
reproduce exactly $R\thicksim 1\div 10kpc$. In general, we can evaluate the
number of nucleons in a self-gravitating system as 
\begin{equation}
N=M/m_{n},
\end{equation}
where $N$ is the number of nucleons of mass $m_{n}$ into self-gravitating
system of total mass $M$\footnote{%
In the present analysis the mass difference between proton and neutron is
not relevant such as it will be shown below. The mass of nucleons is much
larger than the mass of electrons, $m_{p}=1836m_{e}$; therefore we can
neglect the mass of electrons.}. Then, we obtain the relevant results
recalled in Table 2. In the second column the number of evaluated nucleons
is shown, while we find the expected radius of self-gravitating system in
the last column. 
\begin{eqnarray*}
&&\frame{$
\begin{array}{lll}
\text{\textbf{Sys Type}} & \text{\textbf{N.of Nucleons}} & \text{\textbf{%
Eval. Length}} \\ 
\text{Glob. Clusters} & N_{G}\thicksim 10^{63\div 64} & R_{GC}\thicksim
1\div 10pc \\ 
\text{Galaxies} & N_{G}\thicksim 10^{68} & R_{G}\thicksim 1\div 10kpc \\ 
\text{Cluster of gal.} & N_{CG}\thicksim 10^{72} & R_{CG}\thicksim 1h^{-1}Mpc
\\ 
\text{Superc. of gal} & N_{SCG}\thicksim 10^{73} & R_{SCG}\thicksim 10\div
100h^{-1}Mpc
\end{array}
$} \\
\text{Table 2} &:&\text{ Evaluated Length for different self-gravitating
systems}
\end{eqnarray*}
By comparing the last column in Table 2 with the second column of Table 1,
we see a full agreement between the observed and theoretical values of the
radius. It is obvious that if we have only one constituent (e.g. $N=1$),
like a proton or electron, the relation (2) is the standard and well--known
Compton wavelength. Moreover, we have introduced the quantity $h/Mc$ which
is a type of Compton wavelength: 
\begin{equation}
\lambda_{M} =\frac{h}{Mc}.
\end{equation}
It is interesting to note that at astrophysical scale we find a particular
length. This allows us to obtain the exact radius of self-gravitation just
by multiplying with the power of the number of nucleons which are present in
the systems. We can make the following hypothesis: the observed universe
appears self-similar to its quantum constituents. An invariant scale
relation, from the quantum lengths to the astrophysical ones, plays a
fundamental role. As macroscopic system, our universe shows a sort of
quantum and relativistic memory of its primordial phase. The choice to start
with $\alpha =3/2$ is suggested by statistical mechanics. Eq.(2) is strictly
equivalent to 
\begin{equation}
R(N)=l\sqrt{N},  \tag{2'}
\end{equation}
where $l=h/m_{n}c$. Relation (2') is the well--known Random Walk equation or
Brownian motion relation and it was firstly used by Eddington \cite
{Elnaschie3}, \cite{Sidharth},\cite{capozziello}.

In what follows we can observe that $\alpha =3/2$ is a too rough estimation
if other interactions, in addition to gravity, are relevant. For this
reason, we will consider stochastic self-similar processes at atomic scale.
These processes generalize the classic ones. It is shown below that the
nucleus scale is governed by a law like (2) but with a more complicated $%
l=l(N)$.

\section{Classic and stochastic self-similar random process}

Let $\Re $ be real space and $\gamma _{r}\in \Re ^{+}$, then we define a
self-similar (ss) random process for every $r>0,$ 
\begin{equation}
X(s)\overset{d}{=}\gamma _{r}X(rs),\qquad with\qquad s\in \Re ,
\end{equation}
where \ $\overset{d}{=}$\ denotes equality as distributions \cite{Vervaat}.

The relation (5) is invariant under the group of positive affine
transformations, 
\begin{equation}
X\rightarrow \gamma X,\qquad s\rightarrow rs,\qquad \gamma _{r}>0.
\end{equation}
Since\ $\gamma _{r}$\ satisfies the properties 
\begin{gather}
\gamma _{r_{1}r_{2}}=\gamma _{r_{1}}\gamma _{r_{2}},\qquad \forall
r_{1},r_{2}>0, \\
\gamma _{1}=1,  \notag
\end{gather}
then it must have the form 
\begin{equation}
\gamma _{r}=r^{-\delta },\qquad with\qquad \delta \in \Re .
\end{equation}
Thanks to (8) the relation (5) becomes 
\begin{equation}
X(s)\overset{d}{=}r^{-\delta }X(rs),\qquad with\qquad s\in \Re .
\end{equation}
When a process satisfies (5) or (9), it is said to be self-similar or $%
\delta -$self-similar.

A generalization of self-similar random process is obtained by replacing the
deterministic scaling factor $\gamma _{r}=r^{-\delta }$ in (5) or (9) with a
random variable $\widetilde{\gamma _{r}}\in \Re _{0}^{+}$. This variable is
independent of the process to which such a variable is multiplied. Then
eq.(9) becomes 
\begin{equation}
X(s)\overset{d}{=}\widetilde{\gamma _{r}}X(rs),\qquad with\qquad s\in \Re .
\end{equation}
D.Veneziano demonstrated in \cite{Veneziano} that $\widetilde{\gamma_{r}}$
can also be written as $\gamma _{r}=r^{-\widetilde{\delta }}$ with $%
\widetilde{\delta }$ real random variable. Then, these kinds of processes,
called stochastic self-similar (sss) random processes and the previous ones
(ss), can be treated in the same theory. Gupta and Waymire showed that for $%
0<r\leq 1$ the sss processes are dilations, while for $r>1$ the sss
processes are contractions \cite{Gupta},\cite{Gupta2}.

In \cite{Veneziano} the author proved the following relevant theorem: if $%
\widetilde{\delta_{r_{1}}}\overset{d}{=}\widetilde{\delta_{r_{2}}}$ for some 
$r_{1}\neq r_{2},$ then $\widetilde{\delta }$ must be a deterministic
constant $\delta .$ Then, one can treat ss and sss random processes in a
unique scheme. \newline
Moreover, the author gives many relevant properties and generalizations to a 
$d$-dimensional space in the same paper, but we are not going to consider
these properties because they do not fit the objectives of our paper (for
more details see \cite{Veneziano}).

Presently it appears clear there is an agreement between (5), (9) and (2),
(2'). In fact, by defining the deterministic scaling parameter $\gamma
_{r}=r^{-1/2}$, we find 
\begin{equation}
R(N)=\gamma _{r}R(rN);
\end{equation}
then our studies will explore a $1/2$-self-similar random aggregation
process.

By considering electromagnetic and nuclear interactions, relation (9) becomes

\begin{equation}
R(N)=\widetilde{\gamma _{r}}R(rN),
\end{equation}
with $\widetilde{\gamma _{r}}$\ a random variable. In principle, we have to
expect a change from a deterministic scaling parameter $(\gamma )$\ \ to a
random one $(\widetilde{\gamma })$, due to quantum treatment of nuclear and
electromagnetic interactions.

\section{Exact determination of the power law coefficient}

In this section, we invert the present point of view. We consider the mass
and the radius of objects as known quantities and evaluate the power law
respect to the observed data. We discover the validity of power law at all
scales and not only at astrophysical scales. In other words, we obtain the
impressive result that (2) is also correct at solar system, planet, human,
organic cell and nucleon scales. Naturally, we have to expect a small
difference in our model in respect to (2) due to the role of other physical
interactions.

Let us consider the relation 
\begin{equation}
R(N)=\frac{h}{Mc}N^{x},
\end{equation}
where $x$ is the quantity to be determined.

Then, we obtain 
\begin{equation}
x=\frac{\ln (RM/\alpha )}{\ln (N)},
\end{equation}
where $\alpha =h/c=2.2102209\times 10^{-42}Js^{2}m^{-1}$. For the following
evaluation,we are considering 
\begin{align*}
1 pc& =3.085677587\times 10^{16}m, \\
M_{\odot }& =1.98892\times 10^{30}kg, \\
m_{p}& =1.6726231\times 10^{-27}kg \\
h& =6.6260755\times 10^{-34}Js \\
c& =2.99792458\times 10^{8}ms^{-1}.
\end{align*}
which are the well--known values reported in \cite{PhysRev}.

Table 3 summarizes the results in respect to the objects in the length range 
$10pc<R<100h^{-1}Mpc$ and with a mass in the range $10^{6\div 7}M_{\odot
}<M<10^{17}h^{-1}M_{\odot }$. In particular, by following the data in table
1, we obtain: 
\begin{eqnarray*}
&&\frame{$
\begin{array}{ll}
\text{\textbf{System Type}} & x \\ 
\text{Globular Clusters} & x_{GC}=1.5052\div 1.5084 \\ 
\text{Galaxies (Giant)} & x_{G}=1.4975\div 1.5273 \\ 
\text{Galaxies (Dwarf)} & x_{G}=1.5185\div 1.5435 \\ 
\text{Cluster of galaxies} & x_{CG}=1.5185 \\ 
\text{Supercluster of galaxies} & x_{SCG}=1.5180\div 1.5462
\end{array}
$} \\
&&\text{Table 3: Evaluated values of coefficient $x$ in power law} \\
&&\text{for astrophysical objects.}
\end{eqnarray*}
From Table 3, we note that, in the first approximation, $x\simeq 1.5\simeq
3/2.$ This is also in a full agreement with the Fibonacci numbers and the
Golden Mean trough the relation\footnote{%
The connection was suggested by M.S.El Naschie in various pubblications \cite
{Elnaschie5}.} $x=1+\phi $ with $\phi =\frac{\sqrt{5}-1}{2}$$\ $.

If we make the hypothesis that relation (13) is a universal law, then it has
to be real at all scales. Due to other interactions (e.g electromagnetic and
nuclear) the coefficient $x\thicksim 1.5$ and not exactly $1.5$. Table 4
summarizes the results in respect to solar system objects, while Table 5 is
linked to organic matter and finally Table 6 is in respect to the nucleus of
the periodic table of elements. 
\begin{eqnarray*}
&&\frame{$
\begin{array}{ccccc}
\text{\textbf{Object}} & \text{\textbf{Radius(10}}^{6}\text{\textbf{m)}} & 
\text{\textbf{Mass(Kg)}} & \text{\textbf{N}} & x \\ 
\text{Sun} & 6.96\times 10^{2} & M_{\odot } & 1.1892\times 10^{57} & 1.4156
\\ 
\text{Mercury} & 2.439 & 3.2868\times 10^{23} & 1.9650\times 10^{50} & 1.4228
\\ 
\text{Venus} & 6.052 & 4.8704\times 10^{24} & 2.9112\times 10^{51} & 1.4209
\\ 
\text{Earth} & 6.378 & 5.976\times 10^{24} & 3.5728\times 10^{51} & 1.4206
\\ 
\text{Mars} & 3.3935 & 6.3943\times 10^{23} & 3.8229\times 10^{50} & 1.4233
\\ 
\text{Jupiter} & 71.4 & 1.8997\times 10^{27} & 1.1358\times 10^{54} & 1.4205
\\ 
\text{Saturn} & 59.65 & 5.6870\times 10^{26} & 3.4000\times 10^{53} & 1.4232
\\ 
\text{Uranus} & 25.6 & 8.6652\times 10^{25} & 5.1806\times 10^{52} & 1.4228
\\ 
\text{Neptune} & 24.75 & 1.0279\times 10^{26} & 6.1453\times 10^{52} & 1.4219
\\ 
\text{Pluto} & 1.1450 & 1.7928\times 10^{22} & 1.0718\times 10^{49} & 1.4270
\\ 
\text{Moon} & 1.738 & 7.3505\times 10^{22} & 4.3946\times 10^{49} & 1.4254
\end{array}
$} \\
&&\text{Table 4: Calculated values of coefficient $x$ for solar system
objects.}
\end{eqnarray*}
By considering Table 4 we note the impressive constancy of $x\thicksim 1.4$
for the planets of the solar system. The discrepancy of $0.1$, in respect to
the expected value $\alpha =1.5$, could be an effect of the planets not
being a self--gravitating system. For the Sun this discrepancy is a little
bit worse than planets, probably due to not being a self--gravitating system
and because of the effects of nuclear interactions in the interior of the
Sun. The Moon, on the other hand, does not show a discrepancy worse than
planets. As we know, the Moon is a second order system, not
self--gravitating (e.g. there is a rotation around the Earth and another one
around the Sun), but we found this is not relevant in the present study. In
a recent paper, Lynden-Bell and Dwyer have derived from first physical
principles a universal mass-radius relation for planets, white dwarfs and
neutron stars \cite{Lynden}. In the roughest approximation, the proposed
mass-radius relation for planets reduces to $R\sim a_{0}\left(
M/m_{p}\right) ^{1/3}$ where $a_{0}=$\textit{%
h\hskip-.2em\llap{\protect\rule[1.1ex]{.325em}{.1ex}}\hskip.2em%
}$/m_{e}e^{2}$ which is equivalent to (2) when $\alpha =4/3.$ Then, at this
scale, a stochastic process is more appropriate than a self-similar one (as
we will see in (16')). The aim of the present paper is not to obtain an
accurate evaluation of the parameter of the solar system, but just to give a
compatibility of our model with the solar system. A similar approach was
recently presented in \cite{Agop}; also in this work the results reflect the
Cantorian-fractal structure of the space-time. A detailed and interesting
study on the quantization of the solar system was made by L.Nottale,
G.Schumacher and J.Gay \cite{Nottale5}.

At this point it is natural to test the validity of the hypothesis at
organic scale. Table 5 summarizes the analysis on prokaryotic and eukaryotic
cell and human organic matter. In respect to human scale, we have to take
into account that a normal man is taller than he is wide; then we have to
calculate an equivalent radius $R$. This radius is evaluated by considering
the volume of a man as an equivalent volume on a sphere. In order to obtain
the previous result we make the following considerations. Let us define 
\begin{equation*}
\begin{array}{ll}
s & =\text{size} \\ 
cfr & =\text{circumference at half height}=\text{2}\times s \\ 
h & =\text{height} \\ 
R^{\prime } & =\text{radius of }cfr \\ 
V^{\prime } & =\text{volume}=\pi \left( R^{\prime }\right) ^{2}h \\ 
V & =\text{volume of the equivalent sphere}=4/3\pi R^{3} \\ 
R & =\text{radius of the equivalent sphere}=\sqrt[3]{3/4(R^{\prime })^{2}h}
\end{array}
\end{equation*}
where the value $R$ is obtained thanks to the condition $V\prime =V$. 
\begin{eqnarray*}
&&\frame{$
\begin{array}{ccccc}
\text{Organic Mat.} & \text{Radius\textbf{(m)}} & \text{Mass(kg)} & \text{N}
& x \\ 
\text{Proc. cell} & 10^{-6\,\div \,-5} & 10^{-9\,\div \,-8} & 5.9786\times
10^{17\div 18} & 1.4729\div 1.557 \\ 
\text{Euc. cell} & 10^{-5\,\div \,-4} & 1\div 4\times 10^{-7} & 2.3914\times
(2\times 10^{19}\div 10^{20}) & 1.4847\div 1.5557 \\ 
\text{Man} & 0.3039 & 60 & 3.5872\times 10^{28} & 1.5030
\end{array}
$} \\
&&\text{Table 5: Calculated values of coefficient x for organic matter,
where Prc. and Euc.} \\
&&\text{stand for prokaryotic and eukaryotic.}
\end{eqnarray*}
To evaluate the last row of the Table 5 we have considered the following
parameters: 
\begin{equation*}
\begin{array}{ll}
s & =48\text{ (Italian size)} \\ 
cfr & =0.96\ m \\ 
h & =1.60\ m \\ 
R^{\prime } & =0.153\ m \\ 
R & =\sqrt[3]{3/4(R^{\prime })^{2}h}=0.3039\ m
\end{array}
\end{equation*}
Probably not everyone will agree with our previous choice. However, if we
consider the case $R=h$ or $R=R\prime $, the $x$--values result in $x=1.53$
and $x=1.49$ respectively. In other words, our relation can give us the
expected quantities also in organic matter. We can apply relation (2) in
different ways to obtain relevant information about a fixed structure. For
example, if we know the mass of an object we should be able to evaluate its
radius, or when the radius is known we could evaluate its mass. It could be
a good approach to get the dark matter in a structure by just considering
fundamental quantities such as the Plank constant or the speed of light.

To conclude the analysis we have tested the hypothesis on the periodic table
of the elements. The Universe, at all scale, is made with elements of the
periodic table, then we have to determine the value of $x$ for all 103
elements. It is important to take into account the main part of the mass is
in the nucleus; so, we have to consider this scale for estimating $R$. From
nuclear physics we know \cite{Greiner} that 
\begin{equation}
R(A)=R_{0}A^{1/3},
\end{equation}
where $R_{0}=1.1\div 1.5\times 10^{-15}$ m and $A$ is atomic mass. Table 6
summarizes the obtained results \footnote{%
\ We have performed the analysis on all 103 elements, but only the first 26
elements (from H to Fe) are reported here and we jump to the last one, i.e.
Laurentium-Lr.}. We checked to see if the freedom of the choice of $R_{0}$
in the range $1.1\div 1.5$ is relevant and we found it is not; so we give
only one value for $R$ in the table. \newline
As a comment on Table 6, we can see that the value of the power law of 
\textit{He} is the same value obtained for the planets of the solar system
and the Sun. This is obvious! After the nucleosynthesis the ratio [n]/[p] is
frozen (e.g. for $t\sim 1\div 3$ min neutrons and protons link to produce
deuterium nuclei, which create $^{4}He$ thanks to their fusion)\cite{Kolb}.
The fraction of primordial $He$ in the aggregated matter must be the same.
For these reasons the expectation of a value of $x$ for the planets and the
Sun near $He$ is natural. It is interesting to observe that $R_{0}\simeq
h/m_{n}c=1.32\times 10^{-15}$ m. Relation (15) has a simple explanation: in
a given nucleus, Heisenberg's relation of uncertainty implies that a nucleon
occupies a volume $\sim (\Delta r)^{3}$ where $\Delta r\sim h/m_{n}c$ is of
the order of Compton length of a nucleon. Consequently, the nucleus size
must grow like 
\begin{equation}
R(N)=\frac{h}{Mc}N^{4/3}=\frac{h}{m_{n}c}N^{1/3}.  \tag{16}
\end{equation}
By using an sss process we obtain 
\begin{equation}
R(N)=\widetilde{\lambda _{M}}N^{\alpha },\qquad with\qquad \alpha =3/2, 
\tag{16'}
\end{equation}
where $\widetilde{\lambda _{M}}$ \ is the random function of the sss process
with a formal expression $\widetilde{\lambda _{M}}=\frac{h}{Mc}N^{-1/6}$. In
other words, when other interactions act on a system and they are comparable
or more relevant than the gravitational one, the aggregation process makes a
transition from the ordinary self--similar process to the stochastic
self-similar process.

Relation (16') is also in closer agreement with (2) and (15). A relevant
point is the discrepancy of $\sim 0.08\div 0.15$ among the elements in the
periodic table with a value of $x\sim 1.\,\allowbreak 35\div 1.\,\allowbreak
42$ and the theoretical value $\alpha =1.5$. In principle, we can consider
the discrepancy as a consequence of the approximation $m_{n}=m_{p}$. This
approximation is not relevant; in the worst case, that is, $Lr$, we have $%
A-Z=260-103=157$ neutrons. By considering the difference $%
m_{n}-m_{p}=(939.56563-938.27231)$ MeV/c$^{2}=\allowbreak 1.\,\allowbreak
293\,3\times 10^{6}$ eV/c$^{2}=2.\,\allowbreak 305\,5\times 10^{-30}$kg. In
the case of Lr the correct mass becomes $M\prime =M_{0}\ast
103+(M_{0}+2.\,\allowbreak 305\,5\times 10^{-30})\ast 157=\allowbreak
431.9\,6\times 10^{-27}$kg; then we have a relative error $\sigma =$ $\frac{%
\left| M-M\prime \right| }{M}=\frac{\left| 4.\,\allowbreak 316\times
10^{-25}-4.\,\allowbreak 319\,6\times 10^{-25}\right| \allowbreak }{%
4.\,\allowbreak 316\times 10^{-25}}$ $=$ $8.\,\allowbreak 341\,1\times
10^{-4}$. There is a correction in the coefficient $x$ of $0.001,$ which is
not relevant in the present analysis. For this reason the approximation $%
m_{n}=m_{p}$ is not a source of the discrepancy. Another possible reason
could be due to the fact that we do not consider the mass of the electron.
As it is known, $m_{p}=1836\,m_{e}$, but also in this case, we find a
relative error $\sigma ^{\prime }=2.\,\allowbreak 152\,8\times 10^{-4}$
which is not significant. We can conclude that the only possible reason for
the discrepancy is found in the presence of the other interactions (e.g.
electromagnetic and nuclear interactions). In other words, the statistical
geometry of the system is modified by other interactions. We find the trend
of the power law coefficient $x$ as a function of the atomic number $Z$ in
Fig. 1, while Fig.3 shows the histogram for the 103 elements of the periodic
table. From Fig.1 it clearly appears that the ss--process is a rough
approximation; in fact the elements are not on a constant line. This fact
suggests a generalized scaling parameter like $\widetilde{\lambda _{M}}$ of
the stochastic self-similar processes. 
\begin{eqnarray*}
&&\frame{$
\begin{array}{cccccc}
\text{Elements} & \text{R(}10^{-15}\text{m)} & \text{A} & \text{Mass(}%
10^{-27}\text{kg)} & \text{N} & x \\ 
H & 1.5039 & 1.0079 & \allowbreak 1.\,\allowbreak 673\,1 & 1 & - \\ 
He & 2.\,\allowbreak 381\,6 & 4.0026 & 6.\,\allowbreak 644\,3 & 4 & 
\allowbreak \allowbreak 1.\,\allowbreak 419\,9 \\ 
Li & 2.\,\allowbreak 861\,3 & 6.9410 & 11.52\,20 & 7 & 1.\,\allowbreak 388\,8
\\ 
Be & 3.\,\allowbreak 121\,5 & 9.0122 & 14.9600 & 9 & 1.\,\allowbreak 388\,4
\\ 
B & 3.\,\allowbreak 316\,7 & 10.8100 & 1\allowbreak 7.94\,50 & 11 & 
1.\,\allowbreak 373\,4 \\ 
C & 3.\,\allowbreak 435\,2 & 12.0110 & 19.93\,80 & 12 & 1.\,\allowbreak
381\,8 \\ 
N & 3.\,\allowbreak 615\,8 & 14.0067 & 23.25\,10 & 14 & 1.\,\allowbreak
378\,7 \\ 
O & 3.\,\allowbreak 779\,7 & 15.9994 & 26.55\,9 & 16 & 1.\,\allowbreak 376\,3
\\ 
F & 4.\,\allowbreak 002\,5 & 18.9984 & 31.53\,7 & 19 & 1.\,\allowbreak 373\,8
\\ 
Ne & 4.\,\allowbreak 083\,7 & 20.179 & 33.49\,7 & 20 & 1.\,\allowbreak 377\,1
\\ 
Na & 4.\,\allowbreak 265\,2 & 22.98977 & 38.16\,3 & 23 & 1.\,\allowbreak
371\,2 \\ 
Mg & 4.\,\allowbreak 345 & 24.305 & 40.34\,6 & 24 & 1.\,\allowbreak 376\,1
\\ 
Al & 4.\,\allowbreak 499 & 26.98154 & 44.78\,9 & 27 & 1.\,\allowbreak 369\,2
\\ 
Si & 4.\,\allowbreak 559\,5 & 28.086 & 46.62\,3 & 28 & 1.\,\allowbreak 370\,3
\\ 
P & 4.\,\allowbreak 710\,7 & 30.97376 & 51.41\,6 & 31 & 1.\,\allowbreak
367\,7 \\ 
S & 4.\,\allowbreak 765\,2 & 32.06 & 53.22 & 32 & 1.\,\allowbreak 368\,4 \\ 
Cl & 4.\,\allowbreak 927\,7 & 35.453 & 58.85\,2 & 35 & 1.\,\allowbreak 371\,7
\\ 
Ar & 5.\,\allowbreak 127\,7 & 39.948 & 66.31\,4 & 40 & 1.\,\allowbreak 365\,2
\\ 
K & 5.\,\allowbreak 091\,1 & 39.098 & 64.90\,3 & 39 & 1.\,\allowbreak 366\,8
\\ 
Ca & 5.\,\allowbreak 133\,3 & 40.08 & 66.53\,3 & 40 & 1.\,\allowbreak 366\,4
\\ 
Sc & 5.\,\allowbreak 333\,6 & 44.9559 & 74.62\,7 & 45 & 1.\,\allowbreak
364\,3 \\ 
Ti & 5.\,\allowbreak 447\,6 & 47.90 & 79.51\,4 & 48 & 1.\,\allowbreak 363\,4
\\ 
V & 5.\,\allowbreak 560\,5 & 50.9414 & 84.56\,3 & 51 & 1.\,\allowbreak 363\,3
\\ 
Cr & 5.\,\allowbreak 598\,6 & 51.996 & 86.31\,3 & 52 & 1.\,\allowbreak 363\,5
\\ 
Mn & 5.\,\allowbreak 702\,3 & 54.9380 & 9\,\allowbreak 1.19\,7 & 55 & 
1.\,\allowbreak 362\,7 \\ 
\begin{array}{c}
Fe \\ 
...
\end{array}
& 
\begin{array}{c}
5.\,\allowbreak 733\,6 \\ 
...
\end{array}
& 
\begin{array}{c}
55.847 \\ 
...
\end{array}
& 
\begin{array}{c}
9\allowbreak 2.70\,6 \\ 
...
\end{array}
& 
\begin{array}{c}
56 \\ 
...
\end{array}
& 
\begin{array}{c}
1.\,\allowbreak 362 \\ 
...
\end{array}
\\ 
Lr & \allowbreak 9.\,\allowbreak 573\,8 & 260 & 4\,\allowbreak 31.6 & 260 & 
1.\,\allowbreak 354\,8
\end{array}
$} \\
&&\text{Table 6: Calculated values of coefficient x for the periodic table
of elements.}
\end{eqnarray*}

\section{Physical and geometrical consequences of the fundamental scale
invariant law}

In the previous section we verified the validity of the (2) or (16') at all
scales. Then, we considered the Compton wavelength expression as a
particular case of a more general relation, one which is true for all
material structures in the Universe. We discovered a fundamental relation
which demonstrates the self-similarity of the Universe \cite{Elnaschie5}. We
can evaluate the radius of a particular structure when the mass is known.
The relations (2) and (16') show a Universe that has memory of its quantum
and relativistic nature at all scales. In this sense, the Plank constant and
the speed of light play a fundamental role in giving a quantum and
relativistic parameterization of the structures. This reveals why the
astrophysical structures and organic matter have their particular lengths 
\cite{Elnaschie5}.

During the last twenty years much attention and strong research programs
have been dedicated to the determination of Dark Matter. This is one of the
most interesting problems of modern astrophysics. The presented scale
invariant law can be used to evaluate the baryonic mass of the Universe.
From (2) (with $\alpha =3/2$ \footnote{%
\ As we have seen this is the best value when gravity is the only relevant
interaction.}.) we have 
\begin{equation}
M_{U}=\left( \frac{Rc}{h}m_{p}^{3/2}\right) ^{2}=3.2841\times 10^{55}\text{
kg,}  \tag{17}
\end{equation}
which corresponds to a number of nucleons of 
\begin{equation}
N_{nucl}=\frac{M_{U}}{m_{p}}=1.9634\times 10^{82}.  \tag{18}
\end{equation}
In the previous evaluation we considered a Universe with $%
R_{U}=6000Mpc=1.8516\times 10^{26}$ m. Therefore, in a very rough
approximation we can consider a spherical Universe where its density results
in 
\begin{equation}
\rho _{U}=\frac{M_{U}}{4/3\pi R_{U}^{3}}=1.\,\allowbreak 219\times 10^{-26}%
\text{ g/cm}^{3}.  \tag{19}
\end{equation}
By introducing the critical density 
\begin{equation}
\rho _{c}=\frac{3H^{2}}{8\pi G}=2\times 10^{-29}h^{2}\text{g/cm}^{3}, 
\tag{20}
\end{equation}

evaluated with $H=100$ $h$ km s$^{-1}$Mpc$^{-1}$(where $0.5<h<1)$, we found $%
\rho _{U}>\rho _{c}$ which indicates a closed universe. The gravitational
interaction is sufficient to reverse the expansion of the Universe into a
contraction. This conclusion does not agree with the present observations.
Moreover, the cosmological density becomes 
\begin{equation*}
\Omega =\frac{\rho _{U}}{\rho _{c}}=\frac{9.\,\allowbreak 880\,6\times
10^{-27}}{2\times 10^{-29}}=494.\,\allowbreak 03
\end{equation*}
which does not work.

Clearly the approximation of a spherical Universe is too rough. The self
similarity of relation (2), and the exponent equal to 3/2 are the two
fundamental ingredients of fractal geometry. The scale invariant law lives
in a fractal domain.

This means that Nature manifests itself trough its relativistic quantum
fractal geometry aspects. The presented law has memory of its quantum origin
through the Plank constant; of its relativistic origin thanks to the speed
of light and of its fractal nature due to the fractal power law of nucleons.
Therefore, the Universe has fractal dimension. Following \cite{Mandelbrot}, 
\cite{Sylos}, \cite{Falconer} we can define the fractal dimension as
following 
\begin{equation}
D=\underset{R\rightarrow \infty }{\lim }\frac{\ln (N(<r))}{\ln (R)}, 
\tag{21}
\end{equation}
where $N(<R)$ is the number of nucleons inside the radius $R$ and $R$ is the
radius of the structure. Thanks to (21), we can estimate the fractal
dimensions of all astrophysical structure and of the Universe too. Table 7
summarizes these results.

\begin{eqnarray*}
&&\frame{$
\begin{array}{cc}
\text{\textbf{System Type}} & D \\ 
\text{Globular Clusters} & 3.61\div 3.66 \\ 
\text{Galaxies (Giant)} & 3.27\div 3.54 \\ 
\text{Galaxies (Dwarf)} & 3.18\div 3.39 \\ 
\text{Clusters of galaxies} & 3.20 \\ 
\text{Superclusters of galaxies} & 2.94\div 3.15 \\ 
\text{\textbf{Universe}} & 3.13
\end{array}
$} \\
&&\text{Table 7: Fractal Dimension of astrophysical objects}
\end{eqnarray*}
From Table 7 it is very interesting to note a relative coincidence of the
fractal dimension of the Universe with the number $\pi$.

Taking into account the result 
\begin{equation}
D=3.1329,  \tag{22}
\end{equation}
it suggests a Universe whose spatial bound permeates the time dimension.

If we also consider the time, then 
\begin{equation}
D^{(4)}=4.1329.  \tag{23}
\end{equation}
Theoretically speaking by assuming the limitation of measurement accuracy
the previous value can be the Hausdorff dimension, found by El Naschie \cite
{Elnaschie5}

\begin{equation}
D^{(4)}\cong \langle Dim~\varepsilon ^{(\infty )}\rangle _{H}=4+\phi^{3}
=4.236067977.  \tag{23'}
\end{equation}

\bigskip It is also interesting to note that $D^{(5)}=5.1329$, which is
connected with the fine structure constant, i.e

\begin{equation}
\left( D^{(5)}\right) ^{3}=\left( 5.1329\right) ^{3}\cong \alpha _{0}, 
\tag{24}
\end{equation}

as determined by El Naschie in \cite{Elnaschie5}.

The density of the Universe is: 
\begin{equation}
\rho _{U}^{fractal}=\frac{M_{U}}{4/3\pi (R_{U})^{D}}=2.\,\allowbreak
134\,2\times 10^{-30}\text{g/cm}^{3},  \tag{25}
\end{equation}
which is evaluated in the hypothesis of a spatial pseudo-sphere Universe
(See Fig.3). A similar result can be reach by using a different approach
based on the limit set of Klenian groups \cite{Elnaschie4}. Therefore, $\rho
_{U}<\rho _{c}$ indicates an open universe, i.e. the gravitational
interaction is not sufficient to reverse the expansion of the Universe into
a contraction. This conclusion fully agrees with the present observations 
\cite{Bernardis},\cite{Boomerang}. Moreover, the cosmological density
results in 
\begin{equation*}
\Omega =\frac{\rho _{U}}{\rho _{c}}=\frac{2.\,\allowbreak 134\,2\times
10^{-30}}{2\times 10^{-29}}=0.\,\allowbreak 11\ .
\end{equation*}
We may mention at this point that El Naschie in \cite{Elnaschie5} used the
dimensionless gravity constant $G$ to establish a shanon-like entropy $S(G)=%
\frac{\ln G}{\ln 2}+1=\overline{\alpha }_{e\omega }\simeq 128$ where $%
\overline{\alpha }_{e\omega }$\ is the coupling constant at the
Higgs-Electroweak in order to establish quantum gravity.\ 

\section{Conclusions}

In this paper we have discovered that the Compton wavelength relation is a
particular case of a more general stochastic self-similar law. Our model
allows us to realize an actual segregated universe according to the
observations. Thanks to the relation $R=\lambda _{M}N^{\alpha }$, we have a
link between the actual Universe, as observed, and its primordial phase,
when quantum and relativistic laws were in comparison with gravity.

Relation (2) appears interesting not only because it allows us to obtain the
exact dimensions of self-gravitating systems, but it is scale invariant.
Moreover, it is extraordinary to obtain a single expression (see (16'))
linking micro- and macro-universe in such a simple fashion. We have seen
that this law is also valid for organic matter. Finally, the above relation
is also exact in the presence of dark matter where we have more matter and a
larger radius and it can be used again to set the quantities. If we consider 
$\alpha $ from nuclear physics ($\alpha =4/3$), or the value for Hydrogen $%
\alpha _{H}$, we can evaluate the dark matter in the Universe or its related
length. For these reasons it appears more general in respect to the
questions posed in the present paper. There is no breaking point between
microscopic and large scale universe thanks to the validity of (2) (or
better (16')) at particles scale (traditional Compton wavelength). It is
interesting to note that the observations on the large--scale structures and
the Random Walk relation suggest $\alpha =3/2=1.5$ as best value (in
agreement with El Naschie's $E$-infinity Cantorian spacetime, the Golden
Mean and the Fibonacci numbers), while the nuclear physics relation (15) and
\ our Planets scale results suggest $\alpha \sim 4/3=1.33$. Consequently, we
could take into account that $\lambda _{M}$ can also be a random variable $%
\widetilde{\lambda _{M}},$ linked to the other interactions in addition to
the gravitational interaction. Therefore, the self-similarity model has to
be generalized by a stochastic self-similar model. This confirms the
fractality of power law (2) which tends to be a more general theory (that is
the theory of stochastic self-similar processes which we have just mentioned
but the study of these kinds of processes is not the aim of our paper). In a
certain sense, gravity was analyzed as a statistical property of space-time
and the processes in it.

Our conclusion is that the fractal power law suggests a fractal Universe.
Therefore, we can state that nature uses the language of a \textbf{%
relativistic, quantum and fractal geometry}.

\subsubsection*{Acknowledgements}

The authors wish to thank S.Capozziello for suggestions.


\begin{thebibliography}{99}
\bibitem{Sakharov}  A. Sakharov, Zh.Eksp.Teor.Fiz \textbf{49}, 245, 1965.

\bibitem{Nottale1}  L. Nottale, Fractal Space-Time and Microphysics: Towards
a Theory of Scale Relativity, World Scientific, 1993.

\bibitem{Nottale2}  L. Nottale, Chaos, Solitons \& Fractals, \textbf{4},
361, 1994.

\bibitem{Nottale3}  L. Nottale, Chaos, Solitons \& Fractals, \textbf{6},
399-410, 1995.

\bibitem{Nottale4}  L. Nottale, Astron. and Astrophys. \textbf{327},867-889,
1997.

\bibitem{Elnaschie3}  M.S. El Naschie, On the unification of the fundamental
forces and complex time in the $\varepsilon ^{(\infty )}$ space, Chaos,
Solitons and Fractals, \textbf{11}, 1149-1162, 2000.

\bibitem{Sidharth}  B.J.Sidharth, Chaos Solitons Fractals 11, 2155, 2000;
Chaos Solitons Fractals 12, 795, 2001.

\bibitem{capozziello}  S.Capozziello et al., Mod.Phys.Lett. A16, \textbf{693}%
, 2001.

\bibitem{Rees}  M. Rees, Our Cosmic Habitat, London: Weidenfeld and
Nicolson, 2001.

\bibitem{Lapparent}  V.de Lapparent, M.J.Geller and J.P.Hucra,
Astrophys.Jour. \textbf{302}, L1, 1986.\newline
M.J.Geller and J.P.Hucra, Science \textbf{246}, 897, 1989.

\bibitem{La}  D.La and P.J.Steinhardt, Phys.Rev.Lett.\textbf{62}, 376, 1989.

\bibitem{Buch}  I.L.Buchbinder, S.D.Odintsov, and I.L.Shapiro, Effective
Action in Quantum Gravity, IOP Publishing, Bristol, 1992.

\bibitem{Stelle}  K.S.Stelle, Phys.Rev.D \textbf{16}, 953, 1977.
Gen.Rev.Grav.\textbf{9}, 353, 1978.

\bibitem{Will}  C.M.Will, Theory and Experiments in Gravitational Physics,
Cambridge Univ.Press, Cambridge, 1993.

\bibitem{Penrose}  R.Penrose, The Emperor's New Mind, Oxford University
Press, 1989.

\bibitem{Elnaschie1}  M.S. El Naschie, On the uncertainty of Cantorian
geometry and the two slit experiment, Chaos, Solitons and Fractals, \textbf{9%
}(3), 517-529, 1998.

\bibitem{Elnaschie2}  M.S. El Naschie, Penrose universe and Cantorian
spacetime as a model for noncommutative quantum geometry, Chaos, Solitons
and Fractals, \textbf{9}, 931-933, 1998.

\bibitem{Connes}  A. Connes, Noncommutative Geometry, Academic Press, New
York, 1994.

\bibitem{Cook}  T.A. Cook, The curves of life. Originally published by
Constable and Company, London 1914; Reprinted by Dover Publications, New
York. See also: M.S. El Naschie, Multidimensional Cantor-like Sets ergodic
behaviour, Speculation in Science \& Technology, \textbf{15}(2), 138-142,
1992.

\bibitem{Vajda}  S. Vajda, Fibonacci and Lucas Numbers and the Golden
Section, J.Wiley, New York, 1989.

\bibitem{Binney}  J.Binney and S.Tremaine, Galactic Dynamics, Princeton
University Press, Princeton, 1987.

\bibitem{Vorontsov}  B.A.Vorontsov-Vel'yaminov, Extragalactic Astronomy,
Harwood Academic Pub, London, 1987.

\bibitem{Abell}  G.O.Abell, Astroph.Jou.S \textbf{3}, 211, 1958.

\bibitem{Peebles}  P.J.E.Peebles, Principles of Physical Cosmology,
Princeton University Press, Princeton, 1993.

\bibitem{Vervaat}  W.Vervaat, Bull.Int.Statist.Inst. \textbf{52}, 199, 1987.

\bibitem{Veneziano}  D.Veneziano, Fractals, Vol.7, No.1, 59, 1999.

\bibitem{Gupta}  V.K.Gupta and E.C.Waymire, J.Geophys. Res. \textbf{95},
1999, 1990.

\bibitem{Gupta2}  V.K.Gupta et al., Water Resour.Res.\textbf{95}, 3405, 1994.

\bibitem{PhysRev}  Phys.Rev.D, Vol.54, N.1, Part 1, 1996; D.E. Groom et al.,
The European Physical Journal \textbf{C15}, 2000.

\bibitem{Lynden}  D.Lynden-Bell and J.P.O'Dwyer, One mass-radiusrelation for
Planets, White Dwarfs and Neutron Stars,astro-ph/0104450.

\bibitem{Agop}  M.Agop et al., $\varepsilon ^{(\infty )}$ Cantorian
space-time, polarization gravitational field and van der Waals-type forces,
Chaos, Solitons and Fractals, \textbf{18}, 1-16, 2003.

\bibitem{Nottale5}  L. Nottale, G.Schumacher, and J.Gay, Astron. and
Astrophys. \textbf{322},1018-1025, 1997.

\bibitem{Greiner}  W.Greiner and J.A.Maruhn, Nuclear Models, Springer
Verlag, 1995.

\bibitem{Kolb}  E.W.Kolb and M.S.Turner, The Early Universe, Addison-Wesley,
NY, 1990.

\bibitem{Mandelbrot}  B. Mandelbrot B, The fractal Geometry of Nature,
Freeman, NY, 1982.

\bibitem{Sylos}  F. Sylos Labini, M.Mountuori, L.Pietronero, Phys.Rep. 291,
1997.

\bibitem{Falconer}  K.J.Falconer, Fractal Geometry, John Wiley and Sons,
1990.

\bibitem{Elnaschie5}  M.S. El Naschie, A review of E infinity theory and the
mass spectrum of high energy partiche physics, to appear in Chaos, Solitons
and Fractals, 2003.

\bibitem{Elnaschie4}  M.S. El Naschie, Klenian groups in $\varepsilon
^{(\infty )}$ and their connection to particle physics and cosmology, Chaos,
Solitons and Fractals, \textbf{16}, 637-649, 2003.

\bibitem{Bernardis}  P.de Bernardis et al., Nature \textbf{404}, 955, 2000.

\bibitem{Boomerang}  S.Masi et al., The BOOMERANG experiment and the
curvature of the Universe, astro-ph/0201137, 2002.
\end{thebibliography}
\end{document}